\documentclass[10pt,a4paper]{article}
\usepackage{pstricks,graphicx,epsfig,color,amssymb,amsmath,amscd}
\begin{document}

\title{Braneworld graviton interactions in early universe phase transitions}
\author{Rizwan U.H.~Ansari $^{\rm a}$, Cedric Delaunay $^{\rm b}$, Rhiannon Gwyn $^{\rm c}$,\\ Anke Knauf $^{\rm c}$, Alexander Sellerholm $^{\rm d}$, Nausheen R.~Shah $^{\rm e,f}$,\\ Federico R.~Urban $^{\rm g,h}$\\ \\
{\small ${\rm ^a}$ School of Physics, University of Hyderabad, Hyderabad-500046, India}\\ 
{\small	${\rm ^b}$ Service de Physique Théorique, CEA Saclay, F91191 Gif-sur-Yvette, France}\\
{\small	${\rm ^c}$ Department of Physics, McGill University, Montr\'eal, QC, H3A 2T8, Canada}\\
{\small	${\rm ^d}$ Department of Physics, Stockholm University, Albanova, SE-106 91 Stockholm, Sweden}\\
{\small	${\rm ^e}$ Enrico Fermi Institute, University of Chicago, Chicago IL 60637, USA}\\
{\small	${\rm ^f}$ HEP Division, Argonne National Laboratory, Argonne, IL 60439, USA}\\
{\small	${\rm ^g}$ INFN, via Saragat 1, I-44100, Ferrara, Italy}\\
{\small	${\rm ^h}$ Department of Physics, University of Ferrara, via Saragat 1, I-44100, Ferrara, Italy}}
\maketitle

\begin{center}
  {\bf ABSTRACT}
\end{center}

\noindent
These are proceedings for the Les Houches Summer School ``Particle Physics and Cosmology: the Fabric of Spacetime,'' held in Les Houches, France, July 31--August 25, 2006. We summarise the discussions of our working group and outline a procedure for calculating gravity corrections to scalar field potentials, as they might be relevant for inflationary physics. We focus on a specific Randall--Sundrum type braneworld scenario and discuss the relevance of Kaluza--Klein gravitons due to the extra warped dimension.

\paragraph{\bf Braneworld Cosmology}
Braneworld models have been suggested to solve the hierarchy problem \cite{rs1,rs2,add,aadd} and have received renewed attention in string--inspired cosmology. One assumes the brane to be a (3+1)--dimensional hypersurface embedded in a higher--dimensional bulk with the standard model fields confined to the brane (and gravity propagating in the bulk). 
Since string theory predicts extra dimensions and contains higher--dimensional objects such as D--branes, it seems natural to assume some underlying string theory motivation, although most braneworld models leave the features of the brane itself quite generic.

We will focus on Randall--Sundrum (RS) type models, which embed one or two branes into a five dimensional AdS bulk. In RS1 \cite{rs1} a static solution of the 5d Einstein equations was derived with two branes of opposite tension at the orbifold fixed points of the finite fifth dimension. The curvature in the AdS bulk leads to a ``warping down'' of physical scales and thus provides a hierarchy between the positive tension ``Planck brane'' and the negative tension ``TeV brane''. Unfortunately, in this setup general relativity is not recovered on the TeV brane (containing our observable universe). What one finds instead is a Brans Dicke theory with negative BD parameter, which is ruled out by observations \cite{bd}.
There are two ways to circumvent this problem: one can either consider the RS2 model \cite{rs2}, in which the negative tension brane is pushed to infinity and our universe is confined to the Planck brane\footnote{In this scenario one faces again the hierarchy problem, but one can add a ``probe brane'' in the infinite AdS bulk \cite{lykran} at precisely such a distance that the effective scale on this probe brane is again the electroweak scale. The only requirement for the probe approximation to be valid is that the probe brane tension be much smaller than the Planck brane tension. This is actually an appealing scenario, because it would allow us to consider cosmology on a TeV brane with arbitrary (albeit small) tension.}; or one could take the stabilisation of the radion (the scalar field associated with the brane distance) into account, which removes the constraint that the brane tensions have to be opposite, but leads to additional, model--dependent terms in the Friedmann equation.

In the first case standard cosmology is recovered at late times \cite{csaki1, jim}. For simplicity we only consider models with  infinite, static fifth dimension (and only gravity in the bulk), but we do not restrict ourselves to the RS fine--tuning condition for the brane tension (i.e. we study an FRW brane with a non--vanishing 4d cosmological constant). The Friedmann equation for this case is \cite{langlois}:
\begin{equation}\label{modfriedmann}
  H^2\,=\,\frac{8\pi G}{3}\,\rho\,\left(1+\frac{\rho}{2\lambda}\right) + \frac{\Lambda_4}{3}-\frac{k_4}{a^2}\,.
\end{equation}
Here, $\rho$ is the energy density on the brane, $\lambda$ is its tension and $G$ is the four--dimensional gravitational constant; $k_4$ is the curvature of the 4d space, and the 4d cosmological constant is given by $\Lambda_4$. This shows that there is a high--energy regime where the dominant contribution to $H^2$ arises from $\rho^2$, whereas the low--energy regime is governed by the usual Friedmann equation $H^2\sim\rho$.

Under the assumption that the 4d metric does not mix with the static fifth dimension, one finds the following relation between 4d and 5d quantities (see e.g. \cite{shiromizu}):
\begin{eqnarray}
  \frac{1}{M_4^2} \,=\, \kappa_4^2 \,=\, \frac{1}{6}\,\lambda\kappa_5^4 \,=\, \frac{\lambda}{6M_5^6}
    \qquad\qquad
  \Lambda_4 \,=\, \frac{1}{2}\,\big(\Lambda_5+\lambda\kappa_4^2\big)\,.
\end{eqnarray}
Only by setting $\Lambda_4 = 0$ does one recover the RS constraint, which is equivalent to $M_5^3 = k M_4^2$, where $k$ is the AdS mass scale and related to the bulk cosmological constant by $\Lambda_5 = -6k^2$.

Standard cosmology can also be recovered in RS1 models \cite{csaki2}, at least in the low--energy limit, once we stabilise the radion \cite{wise, rad2}.
It is then no longer necessary to balance the bulk cosmological constant with appropriate brane tension. However, the high--energy behaviour is unclear and depends on the precise radion potential. We will therefore not elaborate further in this direction, but work with the infinite extra--dimensional model with FRW Planck--brane (or a TeV probe--brane). 

\paragraph{\bf Brane Inflation and the Electroweak Phase Transition}
As previously noted, (\ref{modfriedmann}) has a high--energy regime in which $H\sim \rho/M_5^3$. As an immediate consequence, the dynamics of scalar field driven inflation is modified \cite{linde, roy}.
We want to constrain a particular inflationary model using its predictions for the shape and amplitude of the temperature fluctuations (at the time of the last scattering surface) and comparing them to the WMAP data \cite{wmap}. We will not show the detailed procedure but only quote the main result. In order to have successful 4d inflation on the brane (in the case of chaotic inflation with a potential of the form $V\sim m_\phi^2\phi^2$, where $\phi$ is the usual 4d inflaton field), one requires:
\begin{eqnarray}
m_\phi\approx10^{-4}M_5\,.
\label{inflmass}
\end{eqnarray}

For simplicity, we make a na\"ive estimate of the reheating temperature. This is not precise, but nonetheless the scenario we describe below holds.
Furthermore, a complete theory of (p)reheating is still lacking, and we are thus led to extract the reheating temperature by means of the relation $\Gamma_\phi\sim H$, where $\Gamma_\phi=\alpha m_\phi$ is a constant decay rate for the inflaton; $\alpha$ being its coupling constant with light degrees of freedom. Using this relation and assuming that at the end of the reheating process we are still in the high--energy regime (an assumption readily confirmed for low values of $M_5$), we obtain the approximate reheating temperature
\begin{eqnarray}
  T_{RH}\sim10^3\alpha^{1/4}\,m_\phi\sim0.1\alpha^{1/4}\,M_5\,,
\label{reh}
\end{eqnarray}
where the result (\ref{inflmass}) was used in the second step. This shows that low--scale gravity also leads to low reheating temperatures, and it is possible to obtain temperatures lower than the Electroweak scale. This implies that the Electroweak phase transition can take place before the completion of reheating\footnote{Several issues arise here, and are mainly related to our lack of confidence with the (p)reheating stage. For instance, since the actual thermalisation mechanism is still unclear we don't know whether we have thermal equilibrium and thus whether it makes sense to speak of an equilibrium plasma, or what its temperature is.}. Consequently, in this scenario we should take into account the entropy dilution after the transition (coming from the inflaton decay), the faster expansion rate during the transition, and the impact of these effects on, for example, the generation of the baryon asymmetry via the Electroweak phase transition.

\paragraph{\bf Gravitons}
In this section we discuss how gravitons couple to matter fields on the brane. Since we work in an infinite extra--dimensional model, we have a continuous spectrum of KK graviton states, including the conventional 4d graviton as the zero mode.
Matter couples to gravitational fluctuations $h_{AB}$ via the energy--momentum tensor $T_{AB}$,
\begin{eqnarray}\label{sint}
   S_{\rm int} \,\sim\, \int d^4x\,dy\, h_{AB}\, T^{AB}\,,\quad A,B=0\ldots 4\,,
\end{eqnarray}
where the 5d energy--momentum tensor of matter confined to the brane becomes (in the weak gravitational field limit) 
\begin{equation}
  T_{AB}=\eta^\mu_A \eta^\nu_B T_{\mu\nu}(x)\delta(y_c)
\end{equation}
and $h_{AB}(x,y)$ are the perturbations specified below in (\ref{fluct}).
This means, that after integrating out the fifth dimension $y$ in (\ref{sint}), the effective 4d coupling will be determined by the amplitude of $h_{AB}(x,y)$ at the (Planck or probe) brane position $y_c$. 

We can parametrise the 4dim perturbations\footnote{This is after the usual KK reduction of the 5dim perturbations to a tensor, two vectors and a scalar in 4 dimensions.} following \cite{rs2} as
\begin{eqnarray}\label{fluct}
  ds^2=\left[\left(e^{-2k|y|}\,\eta_{\mu\nu}+h_{\mu\nu}(x,y)\right)dx^\mu dx^\nu\right]-dy^2\,,\quad 
    \mu,\nu=0\ldots 3\,,
\end{eqnarray}
with the factorisation $h(x,y)=e^{ip\cdot x}\psi(y)$. Plugging these perturbations in the 5dim Einstein equations, 
we find that $\psi$ obeys a Schr\"odinger--type equation
\begin{eqnarray}\label{schr}
  \left[-\frac{m^2}{2} - \frac{1}{2}\partial_z^2 +\frac{15 k^2}{8(k|z|+1)} - \frac{3k\delta(z)}{2}\right]\psi(z)=0\,,
\end{eqnarray}
where the coordinate $z$ is related to $y$ by $e^{k|y|}dy=dz$, and we have rescaled the perturbations according to $h(x,y)\rightarrow e^{k|y|}h(x,y)$. The explicit solution to this equation for the zero mode as well as higher KK modes has been worked out in \cite{rs2} (see also \cite{rubakov}). The zero mode shows a strong localisation at the Planck brane, whereas the massive graviton modes are suppressed there and approach asymptotically a plane wave for large $mz$. This means that the zero mode coupling at the Planck brane, given by $\psi(0)$, will be much larger than that of individual massive KK states (their contribution might be significant nevertheless, as we sum over a large number of them). On a TeV probe brane, on the other hand, we find the opposite coupling hierarchy. 

We would like to study the impact of gravitational corrections on phase transitions. We will therefore compute the effective potential for a scalar field that undergoes spontaneous symmetry breaking, like the Higgs. For the Electroweak phase transition (which would be interesting e.g. for baryogenesis models) we would have to couple all standard model fields to the KK gravitons. As a slightly less ambitious step, let us consider the one--loop effective potential of a single scalar field coupled to gravity, which could be relevant for the inflationary era.

The one--loop effective potential is given as a sum over all 1PI diagrams with a single scalar/graviton loop.  This can be re--summed into a one--loop vacuum diagram (a loop with all external legs removed). Our task is therefore to calculate this diagram for the massless graviton as well as the continuum of massive KK states. If we are located on the Planck brane, we should integrate over the full KK mass spectrum (up to a cutoff of the order of Planck mass); if we live instead on a probe brane somewhere in the infinite fifth dimension we only need to take masses $10^{-4}$eV$\le m\le 1$TeV into account \cite{lykran} (assuming a hierarchy such that the fundamental scale on the probe brane is precisely 1TeV). In both cases we have to impose a UV cutoff on the loop integral. This will not yield a normalisable result: even at one loop we will encounter an explicit cutoff dependence. This is due to the well--known fact that gravity cannot be renormalised as a QFT.

The KK decomposition, graviton propagators and their coupling to matter and gauge fields have been considered for toroidal extra dimensions \cite{wells, lhz}; we want to repeat the analysis for the infinite AdS bulk. This is not too difficult, as AdS is conformally related to flat space and we can easily obtain the linearised gravity Lagrangian for this case and couple it to the energy--momentum tensor of a scalar field. We hope to see some non--negligible contribution to the symmetry--breaking potential that could result in interesting consequences for the (order of the) phase transition, inflationary dynamics and baryogenesis \cite{us}.

\paragraph{\bf Acknowledgements}
We would like to thank the organisers and staff of the school for creating a very enjoyable and stimulating atmosphere. We are also grateful to Jim Cline and Osamu Seto for helpful discussions. F.U. is grateful to the HEP group at McGill University, where part of this work was carried out. F.U. is supported by INFN under grant n.10793/05. The work of A.K. is funded by an NSERC grant. A.S. is grateful for support from the Carl--Erik Levin foundation. R.G. is supported in part by a Chalk--Rowles fellowship.

\end{document}